\documentclass[twoside,12pt]{article}

\pdfoutput=1
\usepackage{graphicx}
\usepackage{epsfig}
\usepackage[utf8]{inputenc}
\usepackage{url}
\usepackage[cal=esstix]{mathalfa}
\usepackage{amsmath}
\usepackage{amsthm}
\usepackage{amssymb} 
\usepackage[binary-units=true,detect-all,mode=text]{siunitx}
\usepackage{listings}
\usepackage{float, placeins}
\usepackage{caption, subcaption}
\usepackage{dblfloatfix,fixltx2e}
\usepackage[font={it}]{caption}
\usepackage[margin=2.0cm]{geometry}
\usepackage{cancel}
\usepackage{xfrac}
\usepackage{xspace}
\usepackage{booktabs}
\usepackage[binary-units=true]{siunitx}
\usepackage{accents}
\usepackage{titling}
\usepackage[super]{nth}
\usepackage[toc,page]{appendix}
\usepackage{algorithm}
\usepackage{algpseudocode}
\usepackage{bm} 
\usepackage{authblk}
\usepackage{color}
\definecolor{amethyst}{rgb}{0.6, 0.4, 0.8}

\usepackage[colorlinks=true, linkcolor=blue, citecolor=blue, urlcolor=blue]{hyperref}

\begin{document}

\title{On the Utility Function of Experiments in Fundamental Science}

\author[1,2,3]{Tommaso Dorigo}
\author[4,2]{Michele Doro}
\author[5]{Max Aehle}
\author[5]{Nicolas R.\ Gauger}
\author[1,2,4]{Muhammad Awais}
\author[10]{Rafael Izbicki}
\author[6]{Jan Kieseler}
\author[9]{Luca Masserano}
\author[4,7]{Federico Nardi}
\author[2,4,8]{Luis Recabarren Vergara}

\affil[1]{Luleå University of Technology, Laboratorievägen 14, 97187 Luleå, Sweden}
\affil[2]{INFN - Sezione di Padova, via F. Marzolo 8, 35131 Padova, Italy}
\affil[3]{Universal Scientific Education and Research Network, Italy}
\affil[4]{Universit\`a di Padova, Dipartimento di Fisica e Astronomia ``G.Galilei'', via F. Marzolo 8, 35131 Padova, Italy}
\affil[5]{Chair for Scientific Computing, University of Kaiserslautern-Landau (RPTU), Gottlieb-Daimler-Stra{\ss}e, 67663 Kaiserslautern, Germany}
\affil[6]{Karlsruhe Institute for Technology, Kaiserstra{\ss}e 12, 76131 Karlsruhe, Germany}
\affil[7]{Laboratoire de Physique de Clermont Auvergne, 4 Avenue Blaise Pascal
63170 Aubière, France}
\affil[8]{Centro di Ateneo di Studi e Attivit\`a Spaziali "Giuseppe Colombo", Via Venezia 15, I-35131 Padova, Italy}
\affil[9]{Department of Statistics \& Data Science, Department of Machine Learning, Carnegie Mellon University, Pittsburgh, USA}
\affil[10]{Department of Statistics, Federal University of S\~ao Carlos, S\~ao Carlos, Brazil}
\date{\today}
\maketitle

\begin{abstract}
The majority of experiments in fundamental science today are designed to be multi-purpose: their aim is not simply to measure a single physical quantity or process, but rather to enable increased precision in the measurement of a number of different observable quantities of a natural system, to extend the search for new phenomena, or to exclude a larger phase space of candidate theories. Most of the time, a combination of the above goals is pursued; this breadth of scope adds a layer of complexity to the already demanding task of designing the measurement apparatus in an optimal way, by defining suitable geometries and choosing the most advantageous materials and appropriate detection technologies. The precise definition of a global optimality criterion may then require experimentalists to find a consensus on the relative scientific worth of those goals.

In this work, we discuss the problem of formulating a utility function for multipurpose experiments, as an enabling step to employ artificial intelligence tools to explore the design space and assist humans in finding solutions at the Pareto front.
\end{abstract}



\section{Introduction}

Research in fundamental science has a long history of synergy with computer science developments. Some of the applications required to study subnuclear matter and interactions, to model galaxy formation and interactions, to describe matter distribution in the universe, and to investigate similar complex systems often require access to extensive computing resources, innovative algorithms, and specialized computing hardware. A glaring example of the interplay between fundamental science and high-performance computing comes from the field of lattice Quantum Chromo-Dynamics (QCD). Lattice QCD calculations involve discretizing spacetime into a grid, or lattice, to solve the equations governing the strong interaction between quarks and gluons. The complexity of those calculations grows exponentially with the resolution of the lattice and the physical parameters being studied, such as temperature or quark masses. To tackle these challenges, researchers have been relying on supercomputers equipped with specialized architectures, and developed innovative algorithms for parallel computation. Over the past few decades the growth in computing performance has enabled lattice QCD to turn from sideline research into an oracle capable of providing precise predictions of hadron masses, decay rates, and interaction cross-sections~\cite{lqcdreview}. 

In light of the above, it should come as no surprise that in recent years experiments in particle and astro-particle physics --the 
 focus of this work, from which we draw a few examples--  have extensively adopted deep learning technologies, integrating them in their data analysis workflows~\cite{aiforhepbook}. In parallel, significant efforts have begun to be directed towards the integration of neural networks and other machine learning models into online data acquisition systems and processes~\cite{fpga}. Quantum Computing (QC) developments are also now closely followed by the community of experimental physicists, with a view to identifying specific use cases where that new technology may provide effective solutions to very hard problems that are amenable to be treated by it, through a successful encoding and implementation of their underlying computing tasks~\cite{qchep}. 

What the above landscape of computer science-driven advancements has largely left behind --at striking variance with the general embracing of deep learning that the community of fundamental science researchers has otherwise been displaying-- is the task of optimizing the design of the instruments that collect the data ultimately needed for the extraction of information on the studied phenomena\footnote{Recently the community has been trying to organize efforts toward the co-design of hardware and software of future experiments; examples come from the MODE Collaboration~\cite{MODE,toward,progress,tdprogr} and the new EUCAIF initiative~\cite{eucaif}.}.
What the community has so far been referring to as ``optimization'' of particle detectors and other similar apparatus is typically constituted by the appraisal of a quite limited set of design parameters that modify in non-radical ways a pre-defined, experience-driven design concept through the discrete sampling of their possible values. The crucial hindrance to a more extensive investigation of the design space has in the past been the stochastic nature of the physical reactions at the basis of data-generating processes, which makes the likelihood function intractable~\cite{cranmer}. Further, because of the absence of a quantitative recipe to combine the merits of different properties of the apparatus, an evaluation of the relative worth of different choices is usually performed on a narrow subset of the scientific results achievable by the experiment.
The typical {\em modus operandi} includes the generation of large high-fidelity simulations of the relevant physical phenomena, followed by careful modeling of the resulting interaction of particles with the apparatus. Those data constitute the basis of the evaluation of a limited number of proxies of the overall experimental goals. {\em E.g.}, in the design of a detector for a particle collider targeting precision Higgs physics, those proxies could consist in the statistical uncertainty on the self-couplings of the Higgs boson achievable with a given amount of data; while for a gamma-ray observatory such as the one we discuss in Sec.~\ref{s:swgo} {\em infra}, the considered proxy could be the discovery sensitivity to point sources in the sky as a function of energy. 

It should be self-evident that the discrete sampling of just a handful of design configurations cannot in earnest be considered a full optimization of the detector. The significant limitations connected to the large computing cost of high-fidelity \texttt{GEANT4}~\cite{geant4} or \texttt{CORSIKA}~\cite{corsika} simulations provide a rationale for the procedure but do not justify it any longer today: progress in deep learning techniques and development of automatic differentiation tools are enablers of a completely different approach to the problem, which the community cannot any longer afford to ignore. In particular, the success of today's generative models offers effective bypass solutions to the likelihood intractability.

\begin{figure}[h!]
\begin{center}
\includegraphics[width=12cm]{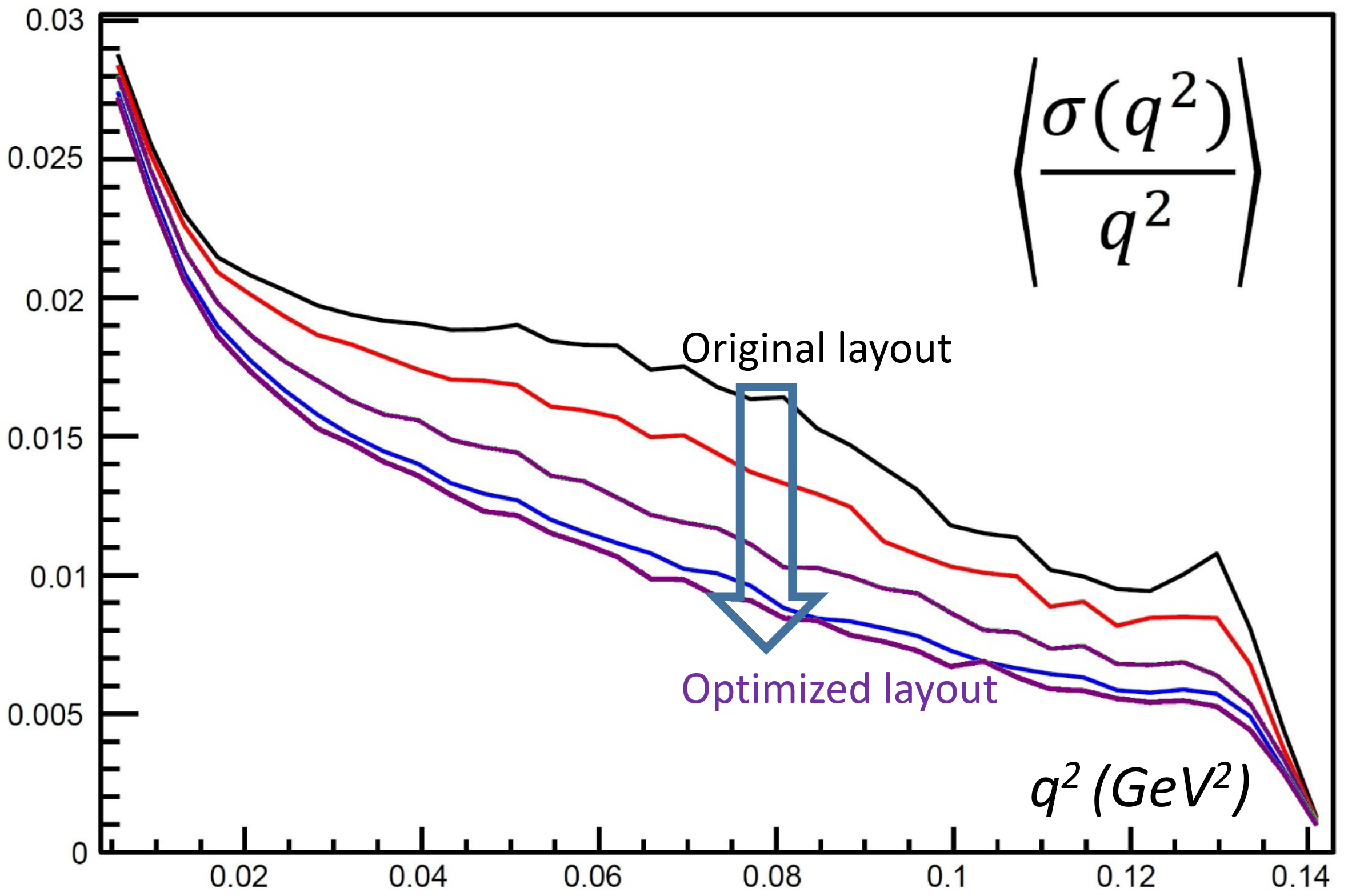}
\caption{Reduction in the estimated relative uncertainty in event $q^2$ as a function of event $q^2$ for elastic muon scattering, for progressively improved layouts of the detection apparatus. Figure reprinted from~\cite{muone}.}
\label{f:muone}
\end{center}
\end{figure}

A motivation for investigating the potential of a more holistic approach to the design of experiments is offered by observing that ``experience-driven'' design choices may actually fall very far from optimality even in the simplest use cases. A proof comes from the study of the relevant figure of merit of MuOnE, an experiment that aims to determine with high precision the differential cross section of elastic scattering of muons off electrons at high energy~\cite{muone}. A reduction of a full factor of two in the relative resolution in event $q^2$, a quantity closely tracking the objective of the intended measurement, was demonstrated by a simple exploration of the geometry of the apparatus, when crucial co-design elements were included in the investigation~\cite{muoneopt} 
(see Fig.~\ref{f:muone}). 

A gain of a factor of two in sensitivity is a unheard-of, off-the-scale performance improvement in particle physics applications, where researchers usually have to have to work hard if they wish to improve their figures of merit by 10 or 20 percent\footnote{A fitting example is the intense and long-term effort spent by the CMS collaboration to precisely measure the Higgs boson decay to photon pairs. Here, machine learning methods were deployed to increase photon energy resolution, obtaining improvements ranging from 5 to 30\%~\cite{higgsggcms}.}. Yet similar potential improvements may be there, waiting to be achieved at no significant increase in cost or complexity even in experiments that are admittedly not at the high end in the complexity scale
of similar endeavors in high-energy physics (HEP). The fact that a factor of two improvement is achievable on top of a design already well thought-over and agreed upon by leading detector experts, as in the case of MuOnE, should count as a valid demonstration of the potentially ground-breaking nature of intensive and specialized machine learning solutions to optimization tasks in experiment design. 

Two factors require to be carefully appraised for their potential as show-stoppers in the way to artificial-intelligence-powered experiment optimization: on one side the high dimensionality of the design space, and on the other the multi-purpose nature of the utility function of typical large-scale experiments. The MuOnE example mentioned {\em supra}, in truth, fails to deliver as representative on both counts. The dimensionality of the parameter space considered by its optimization task lays in the $O(10)$ range. Furthermore, the experiment is uncharacteristically focused toward a single experimental goal: reducing as much as possible the uncertainty in the differential cross section of muon-electron scattering at high scattering $q^2$, as an enabling step toward reducing the leading systematic uncertainty in the theoretical prediction for the muon $g-2$ anomaly, which constitutes the sole reason for building the experiment. 
We briefly comment on the two mentioned factors {\em infra}, before we proceed to examine the latter in more detail in the remainder of this work.

\subsection{The design space}

The complexity of multipurpose detectors and similar instruments employed in fundamental science experimentation
implies that a full investigation of the most effective choices for the materials, the detection technology, and the geometrical arrangement of the system may correspond to the exploration of $O(1,000-10,000)-$dimensional design spaces. Such a large phase space of available options is problematic to explore with discrete sampling or evolutionary optimization techniques. The large dimensionality in question is also multi-faceted and not homogeneous (and thus more complex to explore with automated means), as it stems from different factors, which
have grown in importance in recent years: on one side, new technology for the three-dimensional arrangement of detection elements; and on the other, an increasing breadth of available materials, components, and detection technologies. We can ascribe to the first category, {\em e.g.}, the three-dimensional printing of scintillator elements~\cite{3dprinting}, which enables us to imagine highly-non-trivial arrangements of these devices in the detector volume; similarly, improvements in the manufacturing of silicon pixel detectors have pushed the boundaries in the design options of these devices~\cite{cartiglia}. The second category is contributed by advancements in materials science and manufacturing options, 
and it presents a specific challenge to gradient-based optimization techniques~\cite{toward}
due to the discrete nature of the space of solutions, which does not directly lend itself to continuous scanning. 

Only little more than a decade ago, the simultaneous optimization of $10,000$ parameters, heterogeneous in nature and connected in non-trivial ways to the data-generation mechanisms at the basis of the extraction of information on the studied physical phenomena, would be considered beyond the state-of-the-art of computing tasks. Yet, that feels like long past history today. In profit-driven optimization tasks
the bar is now set several orders of magnitude higher: suffices to say that the training of ChatGPT 4.0
required the simultaneous optimization of an estimated 1.7 trillion parameters. However, that task was significantly expensive --- the sheer cost of electricity required for it is estimated at several hundred million US dollars. Although particle physics is not an inexpensive area of fundamental science, even large experiments are unlikely to be able to invest even a fraction of that sum in R\&D studies any time soon. 

In addition to the large computing costs associated to a thorough exploration of high-dimensional design spaces (and their associated significant carbon footprint), the software tools necessary for that exploration are not off-the-shelf, they 
have to be developed from scratch. Already the very-small-scale end-to-end optimization of MuOnE 
required one person-year of software
development and analysis in order to be carried out; larger and more complex systems may entail development costs that experiments 
could find hard to allocate, especially in an early design stage when funding is usually not yet secured. 

To the rescue in this situation comes the relative non-specific nature of some of the ingredients of the problem. For example, the optimization of a ground-based gamma-ray detector array such as the one we discuss in Sec.~\ref{s:swgo} {\em infra} may require a fast simulation of extended atmospheric showers that may produce a good representation of the secondary particles flux on the ground, given the primary particle identity, energy, and incidence angle. The effort to develop a differentiable model of the secondary particles flux can be significant, but it may be worth spending it if the resulting code is reusable for future optimization tasks of other instruments. Similarly, the development of a differentiable model of the pattern recognition and regression procedures necessary to extract information on the particle flow and incident energy of hadrons in a calorimeter can be a very demanding task, but it becomes an attractive option if the code is designed in a way that ensures flexibility and adaptability to different software pipelines, such that it may be reused to handle detectors of different geometry, components, and materials. In other words, there is an intrinsic modularity
in the software tasks performing the modeling of the data-generating processes, and the reconstruction of the resulting observable features in an instrument; leveraging that modularity makes the construction of differentiable programming models of hardware and software possible, enabling co-design optimization. The MODE Collaboration\footnote {\url{https:\\mode-collaboration.github.io}} has undertaken the ambitious
task of developing reusable and modular software pipelines for gradient-based experiment optimization, and is in the process of building a library of software solutions to the optimization of experimental endeavours of low- to mid-scale complexity. This approach may provide
in the long run the tools necessary to faster solutions to more and more complex optimization questions.

\subsection{Multi-purposedness}

The question that remains on the table is the one of coping with the ambitious nature of large experiments in fundamental science, that usually target not one, but a number of different scientific goals at the same time. From a machine learning perspective, there is a quantum leap in the complexity scale when going from a single, well-defined optimization task (such
as that of the MuOnE experiment we discussed {\em supra}) to multi-target optimization. The reason is connected
to the mathematical specification of the loss function, or more conveniently the negative of the loss, which is sometimes 
called objective function, but which we will rather call {\em utility} in this work. In a multi-objective scenario, the utility function should be written as the sum of different terms, each of which gets multiplied by some constant to 
provide the means to specify our relative appraisal of the different components:\par

\begin{equation}
U = \lambda_1 U_1 + \lambda_2 U_2 + \ldots 
\end{equation}

\noindent
The above definition may induce non-convexity of the total utility function $U$ in the parameter space we wish to explore, even if the two components $U_1$ and $U_2$ are separately convex, when there are interactions between the parameters of the two terms. The set of parameters which provides the maximum of $U$ may also change in a non-linear manner following a modification of the ratio of the $\lambda$ factors. This has two key implications: on one side, the solution becomes highly sensitive to the exact definition we choose for the utility; and on the other side, the maximization task is made considerably more complex, due to the appearance of local minima and saddle points. In such cases, the concept of {\em Pareto front} becomes relevant, as it helps organize combinations of parameter values that provide good utility outcomes using non-dominated sorting. The Pareto front consists of parameter values where any change to improve one component of the utility will necessarily decrease at least one other component. Non-dominated sorting enables the systematic exploration of the Pareto front, identifying suitable compromises between the often conflicting utility components. However, investigating the Pareto front may, in typical situations, be computationally quite expensive.

While Pareto front optimization ups the ante considerably, it is a solvable problem. From a pragmatic point of view, it is a negligible complication when we compare it to the much more problematic task of coming up with an agreed-upon list of utility components to be jointly optimized, and corresponding agreed-upon values of
the $\lambda$ multipliers: one may imagine that large scientific collaborations could be unwilling or unable to pull that off. The main purpose of this work is to examine that problem in some detail, to assess whether it is really a show-stopper in the way of the automated exploration of the configuration space of large experiments.

The present document is structured as follows. In Sec.~\ref{s:hep} we discuss how an experiment-wide utility function can in principle be defined for experiments in particle physics, and in fact we argue how in some cases they are implicitly agreed upon by the researchers
as they define their rules for data collection. In Sec.~\ref{s:swgo} we discuss a use case from particle astrophysics, where we may focus on other distinctive features of similar situations. We offer some concluding observations in Sec.~\ref{s:conclusions}.

\section{Case Study 1: Particle Physics \label{s:hep}}

As a case study centered on experimentation in elementary particle physics, let us focus on the Collider Detector Facility (CDF) experiment at the 
Fermilab Tevatron collider. CDF was commissioned in the early eighties. It started operations with a brief demonstrative data-taking in 1985, then collected data in three main running periods from 1987 to 1988 (Run 0), from 1992 to 1996 (Run 1), and from 2000 to 2012 (Run 2, when the collaboration changed name to ``CDF 2 collaboration'' after a major upgrade). For the purpose of our investigation, we will focus on the first two data-taking runs of the experiment, which collected the data that eventually allowed the collaboration to claim discovery of the top quark~\cite{topdiscovery}. 

\subsection {The CDF experiment}

The CDF experiment is an appropriate testbed to examine how the formulation of an experiment-wide utility function can be expressed in a HEP use case, because of its characteristics and the special role it played within subnuclear physics in the 1990ies. 

CDF was conceived between the end of the 1970s and the start of the 1980s by a small collaboration of American, Italian, and Japanese physicists who set out to build a detector which would study the proton-antiproton collisions produced by the Tevatron collider, a machine which would allow collisions at a center-of-mass energy three times higher than that offered by the CERN $Sp\bar{p}S$ machine, which was by then already in its commissioning phase. It was too late to challenge the CERN facility on the discovery of the $W$ and $Z$ bosons, so the main goal of CDF was the discovery of the last missing quark, top. 

It is interesting to note that already in its early design stage CDF was understood to need a strong solenoidal field to measure with high precision the momentum of charged leptons that were expected from top quark decays; the competitor of CDF, the {D0\hspace{-0.9ex}/} experiment, would just a few years later be designed with the main goal of producing the best determination of the $W$ boson mass, by constructing a high-performance electromagnetic calorimeter which, unhindered from any solenoid material, could achieve high precision in the measurement of electrons energy. This early focus on all-important single goals of the two collaborations (which later both turned out to benefit highly from the multi-purposedness and wide range of measurements and searches enabled by their redundant detection systems and by the Tevatron's unprecedented beam energy) probably has more to do with the competition between the United States and Europe at the HEP frontier than on the perceived worth of different scientific investigations. Still, for the sake of reflecting on the utility function of these experiments, we note that its definition in the early eighties was not really difficult, so that important decisions on the resources allocated to the construction of the detector subsystems could be straightforward to take: for CDF, anything that contributed to catching the top quark took priority. Incidentally, this focused approach made the decision to invest a conspicuous effort in the risky, unprecedented challenge of constructing and operating a silicon microvertex detector (SVX) in the core of CDF an easier one to take~\cite{cdf17}. The SVX turned out to be instrumental in characterizing top pair-production events, through the identification of secondary vertices from $b$-hadron decays.

After the initial engineering run of 1985-86 and the first data-taking period of Run 0 had proven the good
performance of the Tevatron collider, as well as demonstrated the quality of event reconstruction that the CDF detector could achieve, from the outset Run 1 of the Tevatron was expected to deliver a sufficient amount of data and the necessary precision of their reconstruction to warrant success in the discovery of top, if that particle was at reach of the collider energy. But in addition to the top quark, many CDF collaborators were hoping they would manage to make other fundamental, groundbreaking discoveries, by finding evidence of some new physics signal predicted by one of a number of scenarios that extended the Standard Model of particle physics. In fact, despite the incredible successes that the theory of elementary particles had until then achieved, it was generally believed that new phenomena would become accessible by studying physical reactions at the electroweak scale. In particular, some new physics scenarios that extended the Standard Model had just become popular among theorists; most notably, Supersymmetric models~\cite{susy}, which had as a natural mass scale for their particles the one of electroweak symmetry breaking, one which was then fully at reach of the 1.8 TeV collisions that the Tevatron collider was delivering; Technicolor theories~\cite{technicolor} 
were also considered quite promising. 

Further, the SVX was installed in CDF just before the start of Run 1, giving for the first time access to the determination of trajectories of charged tracks with micrometric precision to an experiment operating in the harsh radiation-riddled environment of hadron collisions. Given the very large cross section for production of B hadron states at a proton-antiproton machine, a wealth of entirely new B physics measurements were for the first time at reach of the experimentalists, provided that enough data could be collected by track-based triggers\footnote {Proper secondary vertex-sensitive track triggers were only developed for Run 2 of the Tevatron~\cite{svt}, but CDF in Run 1 could still rely on low-momentum electron and muon triggers for the collection of large data samples enriched with B hadron decays.}. 

Finally, the possibility to test perturbative calculations of quantum chromodynamics (QCD) in a fully new energy regime, and up to several hundred GeV in the momenta of outgoing bodies in two-to-two processes, had spurred a lot of interest in QCD phenomenologists. Diffractive physics was also a very hot topic, and CDF included a small number of experimentalists who 
were highly motivated to shed more light on the complex low-energy phenomena involving the exchange of colourless quanta between the colliding particles. 


To the above {\em pot-pourri} of areas of research where the state of the art of human knowledge could be pushed outwards by new CDF data analyses, one must add the development of new particle detectors for its own sake. A significant number of physicists in CDF were heavily involved in developing improvements in the detection techniques and in the design of new electronic chips to increase the potential of triggering or online data processing. CDF offered to be the ideal testing ground for those new hardware concepts. These ideas need to be included in any consideration of the mix of desiderata that motivated the CDF members in their research.

\subsection{The CDF Trigger System}

At the start of Run 0 (1987), the CDF author list included 227 physicists from 22 institutions in 6 countries; these numbers kept growing during Run 0 and Run 1, and reached 481 from 47 institutions in 12 countries in the year 2002. The experiment 
was led by a pair of spokespersons, whose job included the search for a consensual view in the collaboration on important decisions concerning the operation of the detector and the scientific output of the experiment. They did so by consulting with leaders of several working groups, which included ones dealing with specific hardware subsystems, and ones dealing with the different physics analyses. Of special relevance was a Trigger working group, wherein decisions were taken on the allocation of data collection bandwidth to a number of different rules for the storage of interesting events during the runs of the machine.

During Run 0 and Run 1, the Tevatron yielded proton-antiproton collisions inside the core of the CDF detector at a rate of 280 kHz, yet the data acquisition system could only sustain a rate of about 50-75 events per second that could be fully stored 
on tape. The CDF trigger was the system responsible for online reconstruction and selection that progressively reduced the data rate, allowing for the best possible impedence matching of those two very different numbers. The trigger was therefore a crucial step turning detector readouts into analyzable data on tape, which the experiment relied upon to preserve its discovery and measurement potential.

The trigger system was organized in three levels. Level 1, operated by dedicated hardware boards, used pre-set selection cuts on fast primitives read in from specific detector components (calorimeter towers and muon chambers) to bring the initial inflow of data at the bunch crossing rate of 280 kHz to an output of about 1 kHz. Level 2 was a hybrid of dedicated hardware boards and optimized software algorithms, which could use a more fine-grained information from all detector components to produce a fast reconstruction of the main physics objects present in the event, as a preliminary step to apply a selection which could further reduce the throughput to a few hundred Hz. Level 3, which was run on optimized software in a commercial PC farm, finally performed a speed-optimized reconstruction of the full events using the same algorithms used offline. The application of further selection requirements which could now be based on the measured physics objects brought the throughput further down to about 50-75 Hz, which was a rate at which the data acquisition system could sustain the writing of the full event information, properly reduced by previous processing, into data tapes. It is worth noting how the crucial action of Level 2 and Level 3 was performed in parallel. Level 2 in particular used four buffers acquiring the input data from Level 1; employing a simple queuing logic, these buffers distributed the input load to hardware processors on a ``first come, first serve'' basis. When all four buffers were busy processing events, data from any further input event accepted by Level 1 would be lost. Level 3 processors would also ignore any input from subsequent events when they were all busy. 

The question posed by the large mismatch between input and output data rates could be simply put as how to decide what data to keep and what data to throw away. A trigger menu was devised, to set minimum energy or momentum 
thresholds on interesting reconstructed objects (electrons, muons, jets, missing transverse energy). The sum of the accept rates of the various trigger conditions had to not exceed 50 Hz or so, lest the data acquisition system would drown in its 
own dirty water, and generate dead time. Effectively, the detector would then be blind for a fraction of the time. For example, if an accept signal were given to 80 events per second, the first 50 would get written, yet the remaining 30 would get 
dropped, being overrun by new ones coming in; even if the dropped events were the most valuable to store, their relative value would never get to be appraised. Dead time was a democratic, but highly ineffective way of coping with a higher rate of 
proton-antiproton collisions than the trigger could handle.

The matter was made worse by the fact that, during its operation, the Tevatron collider  always strove --- and nobody would object, as this maximized the discovery potential of the experiments! --- to achieve the highest density of protons and antiprotons in 
 its beams. This caused each bunch crossing to generate a number variable from a few to a few tens of independent proton-antiproton collisions, depending on the run conditions.  Every time a new run would start, its conditions would be different. Perhaps a trigger menu that had 
 worked well the other day, when the Tevatron delivered a lower instantaneous luminosity to CDF, would today generate a huge dead time, because the larger number of collisions produced by the higher luminosity would generate a higher accept rate by the trigger; further, the relationship was non linear, as the pile-up of several low-energy proton-antiproton collisions that individually would not manage to fire the trigger conditions could conspire to collectively make it. The weekly ``trigger meeting'', aptly held in a meeting room adjacent to the one hosting the Level 2 trigger hardware, was the venue where CDF physicists took informed decisions on which thresholds to raise from which trigger condition, in order to keep the dead time below a physiological level of about 10\%. 

To make a concrete example, there were so-called ``high-p$_t$'' electron and muon triggers. These were constituted by sets of conditions that ensured that the experiment would collect events with high-purity charged lepton candidates of high transverse momentum (p$_t$), typically produced by the decay of $W$ or $Z$ bosons. Due to their significant acceptance of $W$ boson decays, those triggers were ideal for collecting events which could be the result of the production of a pair of top quarks. The full functionality of high-p$_t$ electron and muon triggers had therefore to be preserved, hence their thresholds were never tweaked to cope with variations in instantaneous luminosity. This was a relatively easy decision, as the accept rate of the resulting data streams was relatively small, of the order of 1 Hz. But then there were jet triggers: in principle, jets are the commonest thing one can see in a proton-antiproton collision, and their high cross-section poses a direct threat to a fixed-throughput trigger system: so one could accept to collect only a small fraction of events with high-energy jets. However, many CDF scientists would have loved to collect as many jet events 
as possible. Perhaps 20 Hz of jet events could be stored? Well, that could be a sound decision, but then there 
were scientists who wanted to study B physics, which relied on collecting low-momentum electrons and muons. Those, too, 
were relatively frequent. Should CDF collect more low-momentum leptons or more high-energy jets?

In truth, above the problem has been massively simplified, as the trigger menu in CDF included over a hundred different sets of requirements on the energy and type of objects measured by the online reconstruction. In addition, a dynamical prescaling system was in place to automatically reduce or increase the acceptance rate of specific trigger conditions, by discarding a varying fraction of the events accepted by those conditions. Dynamic prescaling was applied to ``expendable'' datasets, whose collection was in other words conditional to the avoidance of dead time. But the bottomline remains clear: deciding what kind of datasets to collect, and how many events to acquire from each, was equivalent to deciding what was the overall goal of the experiment. This means that in a way, the trigger of a limited-bandwidth hadron collider experiment is tantamount to an operative definition of its utility function.

\subsection{A proxy to the scientific value of different analysis targets}

Eventually, CDF did discover the top quark, as it did produce groundbreaking measurement in a number of other measurements and searches. From the vantage point of {\em a posteriori} examination, we have a way to judge on the relative merit of the various physics results that the experiment produced, which correlates with the impact that those measurements had in the science CDF advanced. This is constituted by the number of citations of the publications that the experiment published based on those measurements. 
While such a metric may not be objectively the best one to assess the scientific worth of different results, it has the merit of being simple to define and relatively unbiased, as long as it compares articles from the same experiment.

Examining the 20 most cited works published by the CDF collaboration based on data collected before the start of Run 2 of the Tevatron, we get the following picture for the number of citations as of January 10, 2025 (see Table 1 for more detail):

\begin{enumerate}
\item Observation of top quark production, and related publications~\cite{cdf1,cdf3,cdf4}: {\bf 6840} citations
\item $J/ \psi$ and $\psi(2S)$ production, plus related publications~\cite{cdf6,cdf9,cdf13}: {\bf 1359} citations
\item Detector descriptions~\cite{cdf2,cdf12,cdf17}: {\bf 2026} citations
\item QCD results on jet properties~\cite{cdf5,cdf14,cdf15}:{\bf 1327} citations
\item Measurement of total p-antip cross section~\cite{cdf8,cdf11}: {\bf 869} citations
\item $B_c$ observation and a related publication~\cite{cdf7,cdf17}: {\bf 827} citations
\item Double parton scattering~\cite{cdf10,cdf19}: {\bf 770} citations
\item W boson mass measurement~\cite{cdf18,cdf20}: {\bf 658} citations
\end{enumerate}

\noindent
Leaving aside the three detector-related articles, after observing that the citations they got (14.5\% of the total) give us a scale of the relative importance of detector developments {\em per se}), we are left with the task of aggregating the seven remaining classes of publications together based on the triggers that produced the respective analyses. First of all, the top and $W$ mass publications were mainly produced by analyzing data collected by high-transverse-momentum lepton triggers, and brought CDF authors collectively 6840 citations; the $J/ \psi$ and $b$ 
physics papers, mainly produced through the study of low-p$_t$ lepton triggers, brought 2186 more citations; and the QCD-based measurements, brought in by study of single and multiple jets and minimum bias triggers, a total of 2966 citations. By normalizing these numbers, we may finally assess that the ``relative value" of the high-p$_t$ lepton triggers, low-p$_t$ lepton triggers, and hadronic activity-based triggers in CDF during Run 1 (the period of data taking which produced the publications above discussed) is respectively 
of 57.0\%, 18.2\%, and 24.7\%, with sub-percent statistical uncertainties (albeit admittedly also carrying a much more significant additional error from their ill-defined nature). 

\begin{table*}[h!]
\tiny
\begin{center}
    \begin{tabular}{llccrl}
    Item & Abbreviated title & Year & Category & Citations & Notes\\
    \hline
    1 & Observation of top quark production [...]~\cite{cdf1} & 1995 & 1 & 4288 &\\
    2 & The CDF Detector: an Overview~\cite{cdf2} & 1988 & 0 & 1988 & Detector description\\
    3 & Evidence for top quark production [...] (PRD)~\cite{cdf3} & 1994 & 1 & 1062 & \\
    4 & Evidence for top quark production [...] (PRL)~\cite{cdf4} & 1994 & 1 & 832 & \\
    5 & The Topology of 3-jet events~\cite{cdf5} & 1992 & 3 & 571 & \\
    6 & $J/\psi$ and $\psi(2S)$ production [...]~\cite{cdf6} & 1997 & 2 & 515 & \\
    7 & Observation of the $B_c$ meson [...]~\cite{cdf7} & 1998 & 2 & 469 & \\
    8 & Measurement of the $p \bar{p}$ Total Cross- Section [...]~\cite{cdf8} & 1993 & 3 & 452 & Special run \\
    9 & Inclusive $J/\psi$, $\psi(2S)$ and b-quark production [...]~\cite{cdf9} & 1992 & 2 & 448 & \\
    10& Double parton scattering in $\bar{p}$p collisions [...]~\cite{cdf10} & 1997 & 3 & 448 & \\
    11& Inclusive jet cross section [...]~\cite{cdf11} & 1996 & 3 & 417 & \\
    12& CDF central and end-wall calorimeter~\cite{cdf12} & 1988 & 0 & 403 & Detector description\\
    13& Production of $J/\psi$ mesons from $\chi_c$ meson decays [...]~\cite{cdf13} & 1997 & 2 & 396 & \\
    14& Charged Jet Evolution and the Underlying event [...]~\cite{cdf14} & 2002 & 3 & 395 & Minimum Bias data \\
    15& Transverse Momentum Distributions of Charged Particles [...]~\cite{cdf15} & 1988 & 3 & 361 & Special run/MB data \\
    16& Observation of $B_c$ mesons [...]~\cite{cdf16} & 1998 & 2 & 358 & \\
    17& The Silicon Vertex Detector [...]~\cite{cdf17} & 1994 & 0 & 354 & Detector description\\
    18& A Measurement of the $W$ boson mass~\cite{cdf18} & 1990 & 1 & 336 & \\
    19& Study of four jet events and evidence for double parton [...]~\cite{cdf19} & 1993 & 3 & 322 & \\
    20& A Measurement of the $W$ boson mass [...]~\cite{cdf20} & 1990 & 1 & 322 & \\
    \hline
    \end{tabular}
    \caption{\label{t:cdf} The twenty CDF publications based on Run 0 and Run 1 data which collected the largest numbers of citiations as of January 10 2025. The Category column indicates the type of physics goals addressed by the study: category 0 refers to detector descriptions; category 1 includes top quark searches and measurements, and electroweak studies; category 2 includes studies of heavy-flavoured hadron properties; and category 3 includes QCD-related studies. See the text for detail.}
\end{center}
\end{table*}
\normalsize

A small refinement of the above numbers comes if we observe that three of the publications affering to the broad category of QCD measurements (numbers 8, 14, and 15 in the list shown in Table~\ref{t:cdf}) were produced through the analysis of data collected by special runs, where the experiment could afford to acquire a significant number of so-called ``minimum bias'' events that produced very little detected signals in the detector, and/or when the Tevatron collider was run at a lower energy corresponding to that previously studied by the $Sp\bar{p}S$ collider at CERN. It is an interesting fact {\em per se} that three of the 20 most cited publications by CDF in Run 0 and Run 1 were based on the analysis of those very quick runs, which subtracted an irrelevant amount of integrated luminosity to the total bounty collected by the experiment! The choice of devoting run time to these QCD measurements was never controversial; the above observation not only justifies the special procedure {\em a posteriori}, but it allows us to note how the existence of a facility that sheds light on previously unexplored territory (the highest center-of-mass energy ever achieved in hadron collisions) offers scientific value through the diversity of studies it enables, some of which may not be a major driver of the decision to build the facility in the first place. In any case, by excluding also those three publications we remain with 14 which break down into these fractions of citations: 63.4\% to top and electroweak related analyses, brought in by high-p$_t$ lepton triggers; 20.3\% to b-quark analyses relying on low-transverse-momentum leptons; and 16.3\% to QCD analyses relying on regular jet trigger data. 

\begin{figure*}[h!]
\begin{minipage}{0.49\linewidth}
\includegraphics[width=\linewidth]{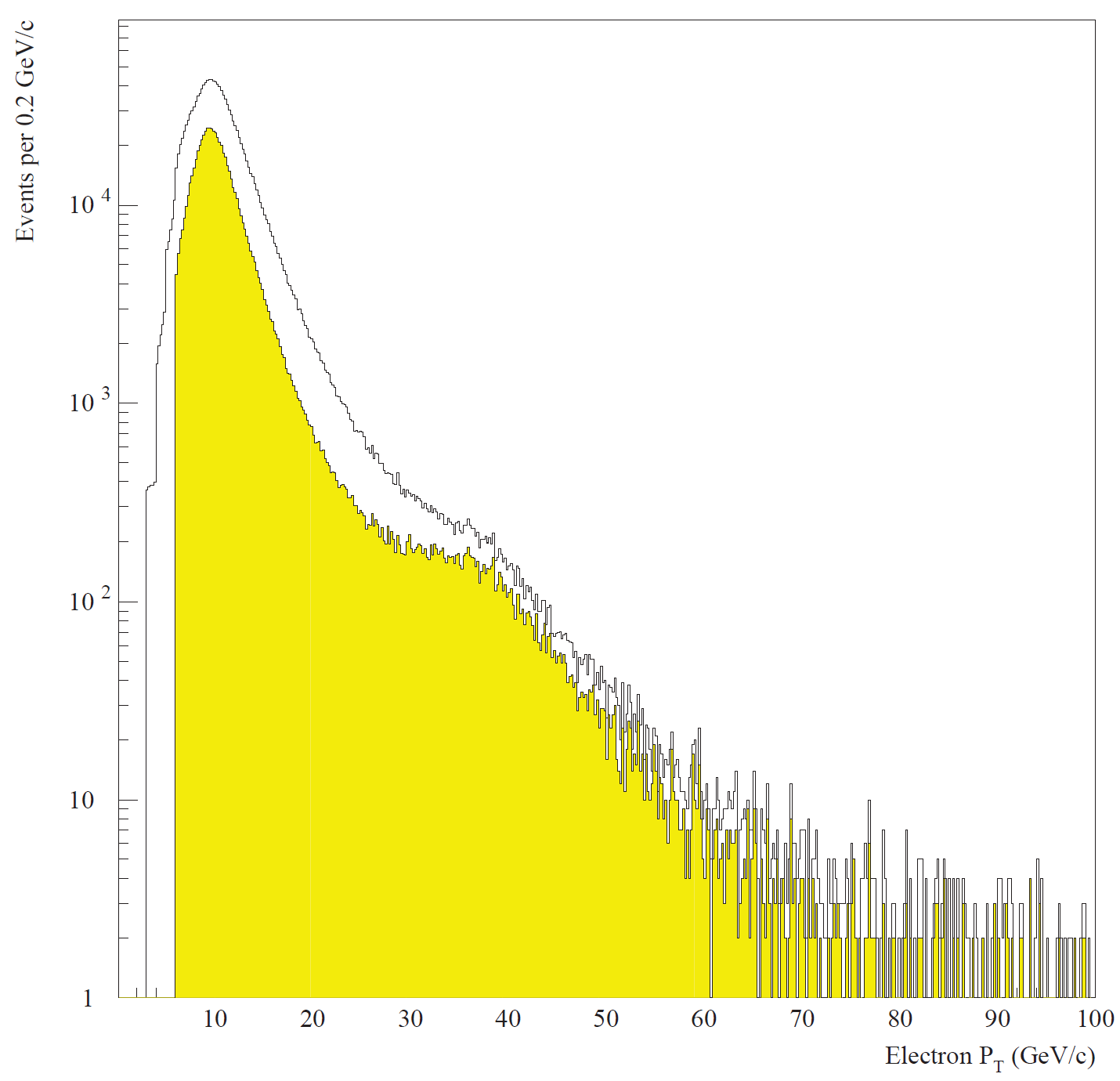}
\end{minipage}
\begin{minipage}{0.49\linewidth}
\includegraphics[width=\linewidth]{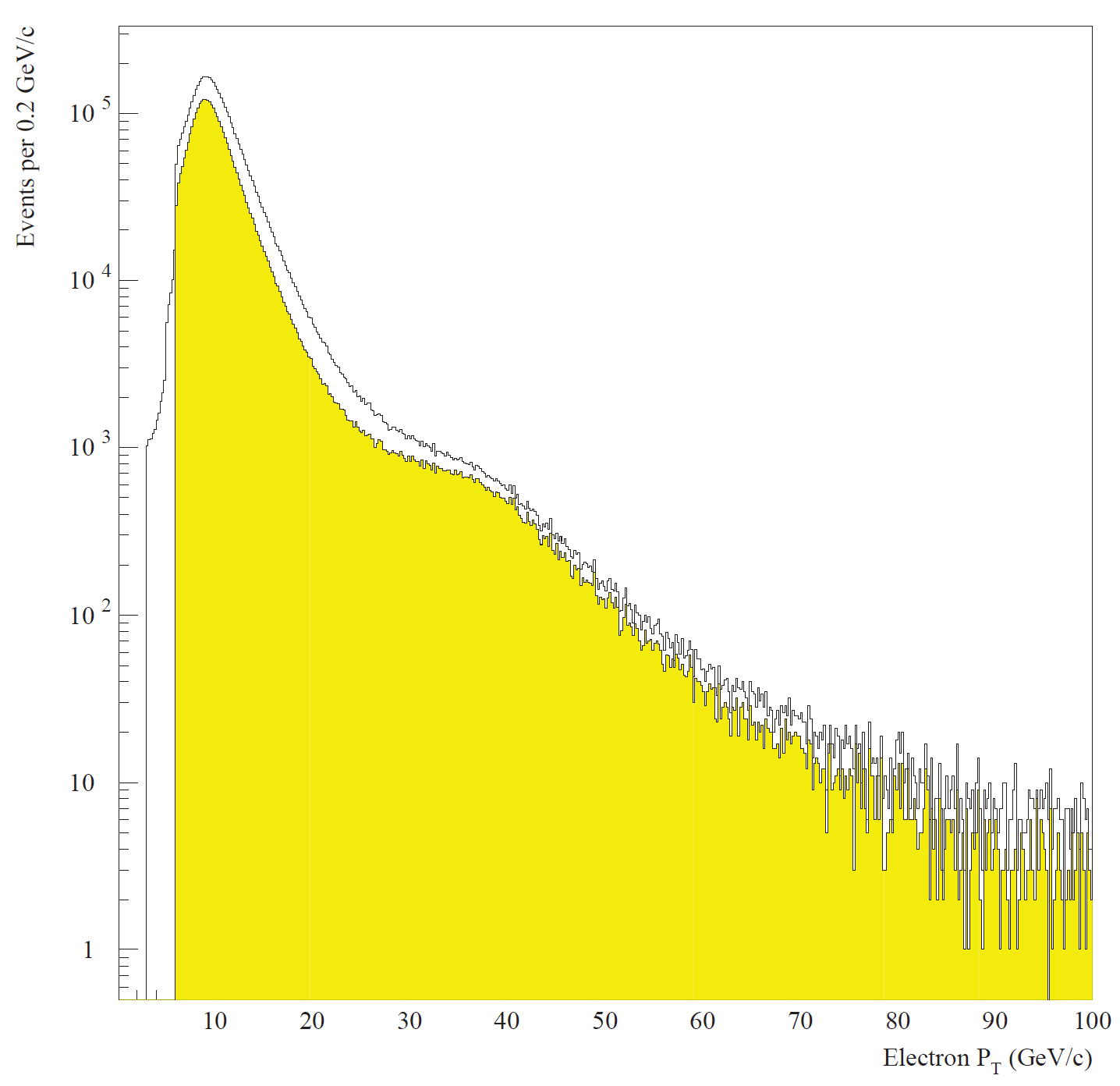}
\end{minipage}
\caption{Transverse momentum distribution of electron candidates collected by inclusive electron triggers in Run 1A (left) and Run 1B (right) before (empty histogram) and after standard quality cuts (yellow histogram)~\cite{cdf4079}. The contribution from $W$ and $Z$ decays is evident in both graphs; it is not significantly affected by the the momentum-dependent selection operated by the trigger, which causes the turn-on shape of the distributions at transverse momenta of about 10 GeV.}
\label{f:inclep}
\end{figure*}

And what were the overall rates of the triggers producing the three classes of events used by the above analysis categories in CDF during Run 1? Alas, those numbers are not easy to determine with precision, as they varied significantly over time; apparently, they were not reported in detail in published documents. Furthermore, the CDF collaboration made the rather upsetting move of not making public their internal CDF notes archive after the experiment stopped its data-taking phase\footnote {Formally, the CDF Collaboration still exists, and a trickle of publications continues to appear on scientific journals to this day, a testimony of the immense value of the data the experiment produced.}. 
What we may note here is the high-p$_t$ lepton triggers took way less than $57.0\%$ of the Level-2 output bandwidth, primarily because of the rarity of the corresponding Level-1 accept rates of data streams containing lepton-rich primitives. Even if the momentum thresholds of the trigger objects of interest (electromagnetic clusters in the calorimeter,
and muon stubs in the muon chambers) as well as the objects' quality criteria had been further loosened from their operation values to try and increase the data statistics, the discovery potential of the collected data would not have changed significantly, as most of the reconstructible top quark events would not require looser thresholds to be acquired (see Fig.~\ref{f:inclep}. In this sense, the bandwidth occupied by top-discovery-related datasets is not a good gauge of the relative value of the resulting physics output by the experiment. At first sight, this also suggests that a top discovery could in principle have been produced by the experiment even if the data acquisition system had been significantly reduced in capacity. However, massive use of low-transverse-momentum lepton triggers as well as of jet triggers was made by the analyses which proved the existence of top quark decays in CDF Run 1 data, because the resulting data samples were instrumental in providing crucial cross-checks~\cite{cdf1,cdf3} and, in the case of jet triggers, a data-driven parametrization of the rate of fake secondary vertex $b$-quark tags in jets~\cite{cdf3}. 
What we can conclude from the above is that even for the single goal of discovering a new particle at a hadron collider, the optimal choice of selection requirements to store collision data is by no means trivial, and it is a problem only amenable by co-design techniques: {\em e.g.}, a larger amount of low-momentum leptons and jet data would have allowed CDF to develop more performant B-tagging algorithms, and to estimate with higher precision their efficiency; but the resulting bandwidth increase of the corresponding triggers would then have had to be balanced carefully, something which can only be done by software tools available since the last decade. 

Analyses which could have benefited from a loosening of the energy, momentum, and quality thresholds of low-transverse-momentum electron and muon candidates clearly include those of heavy-flavored hadrons, which as we showed above brought an overall $20.3\%$ of the citations to prominent CDF Run 1 publications. Here again, the rate of low-transverse momentum electron and muon triggers were much lower than that fraction: {\em e.g.}, the number of events collected in the ``inclusive electron" data stream in Run 1A (1B) amounted to 1,874,289 (6,125,629) events (see Fig.~\ref{f:inclep}), which corresponds to far less than 1\% of the output bandwidth of the experiment during typical Run 1 conditions\footnote{As the integrated luminosity of the mentioned datasets is of $19.4 \pm 0.8$ pb$^{-1}$ and $81.5 \pm 9.0$ pb$^{-1}$~\cite{CDF4320}, the datasets correspond to cross sections in the $70-100$ nb range. This, noted that the instantaneous luminosity during Run 1 averaged to less than $10^{31}$ cm$^{-2}$ s$^{-1}$, indicates a corresponding Level-3 accept rate of the inclusive electron triggers of less than 1 Hz.}; the number of inclusive muon events was even smaller.

Instead, it is much more interesting to observe how QCD triggers, which were almost invariably the ones whose energy thresholds were tweaked upward by experimenters when dead-time had to be reduced, ended up producing a whooping 24.7\% of the high-impact scientific results of CDF in Run 1. Such a datum highlights how the actual scientific impact of a multi-purpose experiment in fundamental physics may turn out to be quite different from the intended one. It also pays justice to those CDF members who indefatigably kept fighting during trigger meetings to avert the prescaling of jet-based trigger selections, or their plain removal from the trigger list\footnote{ In the book ``{\em Anomaly - Collider Physics and the Search for New Phenomena at Fermilab}''~\cite{anomaly}, T. Dorigo describes such a situation in detail.}.

One final word concerns the ``missing in action'' category of CDF articles describing searches for new physics beyond the Standard Model. It is indeed quite striking to observe how no CDF publication of that kind is present among the 20 most cited works. The experiment did invest a significant effort in designing and maintaining triggers that would collect, {\em e.g.}, events with photons, missing transverse energy, isolated tracks, or other non-mainstream trigger primitives that could be sensitive to exotic phenomena; the collective accept rate of the corresponding data streams was quite significant, perhaps of the order of 10 to 20 Hz. This mismatch between investment and return is impossible to avoid, as it rests on the existence or absence of new physics processes; even only one clear beyond-the-Standard-Model signal unearthed by the collaboration would most likely correspond today to the highest-cited publication by CDF. Ultimately, without the benefit of {\em a posteriori}-ness, experiments need to decide how much effort, resources, bandwidth to allocate to fishing expeditions: a degree of arbitrariness in the specification of the utility function is thus impossible to avoid in HEP experiments.

The bottom-line of the above discussion of CDF triggers in Run 1, for what concerns the focus of this work, is that at hadron colliders the limited bandwidth of the data acquisition system demands a quantitative appraisal of the different goals of the experiment; yet the defined working point of the trigger rates and the size of the resulting datasets can only be taken as a qualitative proxy to the overall utility function. 

\subsection {Definition of a multi-target utility at lepton colliders: the BES-III case \label{s:bes3}}

Above we have seen how the definition of an experiment-wide utility function in hadron collider physics experiments may be implicitly related to the issue of reducing the input data flow, which demands crucial decisions in the settings of the trigger system. Other HEP experiments that deal with smaller data rates may not need to take such draconian measures in real-time, but they are often still confronted with defining proxies of an experiment-wide utility in the definition of their data collection schedule. For example, electron-positron colliders of low energy, such as the Beijing Electron Synchrotron (BES) or the Super-KEK-B accelerator in Tsukuba, need to decide how much data to deliver to the detector at different energy points.
The choice directly impacts the precision of different measurements of particle properties, so the run schedule tightly correlates with the relative value given by experimental groups to the different scientific outputs potentially at reach by their detectors.

\begin{table*}[h!]
\tiny
\begin{tabular}{rlccrl}
Item & Abbreviated title & Year & Cit. & L ($fb^{-1}$) & $\sqrt(s)$ (GeV)  \\
\hline
1 & Design and Construction of the BES III Detector~\cite{bes1} & 2010 & 1269 & N/A & N/A \\
2 & Observation of a Charged Charmonium-like Structure in $e^+e^- \to \pi^+\pi^- J/\psi$~\cite{bes2}  & 2013 & 1156 & $0.525$ & $4.260$ \\
3 & Future Physics Programme of BES III~\cite{bes3}& 2020 & 594 & N/A & N/A \\
4 & Observation of a [...] $Z_c (4020)$ and Search for the $Z_c(3900)$~\cite{bes4} & 2013 & 533 & $3.400$ & $3.900-4.420$ \\
5 & Observation of [...] Structure in $e^+e^- \to (D^{*} \bar{D}^{*})^{\pm} \pi^\mp$ at $\sqrt{s}=4.26$GeV~\cite{bes5} & 2014 & 413 & $0.830$ & $4.260$\\
6 & Observation of a Charged $(D\bar{D}^{*})^\pm$ Mass Peak in $e^{+}e^{-} \to \pi D\bar{D}^{*}$ at $\sqrt{s} =$ 4.26 GeV~\cite{bes6} & 2014 & 391  & $0.525$ & $4.260$\\
7 & Measurement of the $e^+ e^ \to \pi^+ \pi^-$ cross section [...] using initial state radiation~\cite{bes7} & 2016 & 341 & $2.930$ & $3.773$ \\
8 & Precise measurement of the $e^+e^-\to \pi^+\pi^- J/\psi$ cross section [...] 3.77 to 4.60 GeV~\cite{bes8}& 2017 & 285 & $9.000$ & $3.770-4.600$ \\
9 & Observation of $e^+e^ \to \gamma X$(3872) at BESIII~\cite{bes9}& 2014 & 251 & N/R & $4.009-4.420$\\
10 & Observation of a Near-Threshold Structure [...] in $e^+e^- \rightarrow K^+(D_s^-D^{*0}+D_s^{*-}D^0$)~\cite{bes10}& 2021 & 233 & $3.700$ & $4.628-4.698$\\
11 & Evidence of Two Resonant Structures in $e^+ e^- \to \pi^+ \pi^- h_c$~\cite{bes11}& 2017 & 205 & $5.600$ & $3.896-4.600$ \\
12 & Precision measurement of the integrated luminosity [...]~\cite{bes12}
 & 2015 & 203 & N/R & $3.097$ \\
13 & Confirmation of [...] and observation of [...] $X(2370)$ in $J/\psi\to \gamma \pi^+\pi^-\eta^\prime$~\cite{bes13}  & 2011 & 200 & N/R & $3.097$\\
14 & Polarization and Entanglement in Baryon-Antibaryon Pair Production [...]~\cite{bes14} & 2019 & 198 & N/R & $3.097$\\
15 & Search for hadronic transition $\chi_{cJ} \to \eta_c\pi^+\pi^-$ and observation of $\chi_{cJ} \to K\overline{K}\pi\pi\pi$~\cite{bes15}& 2013 & 195 & N/R & $3.686$ \\
16 & Measurements of absolute hadronic branching fractions of $\Lambda_{c}^{+}$ baryon~\cite{bes16}
 & 2016 & 191 & $0.567$ & $4.599$ \\
17 & Observation of $Z_c(3900)^{0}$ in $e^+e^-\to\pi^0\pi^0 J/\psi$~\cite{bes17}
 & 2015 & 188 & $2.810$ & $4.190-4.420$\\
18 & Measurement of the integrated luminosities [...] at $\sqrt{s}=$3.650 and 3.773 GeV~\cite{bes18}& 2013 & 184 & $2.917$ & $3.650-3.773$\\
19 & Branching fraction measurements of $\chi_{c0}$ and $\chi_{c2}$ to $\pi^0\pi^0$ and $\eta\eta$~\cite{bes19}& 2010 & 179 & N/R & $3.686$\\
20 & Observation of $e^+e^- \to \pi^0 \pi^0 h_c$ and a Neutral Charmoniumlike Structure $Z_c(4020)^0$~\cite{bes20}& 2014 & 166 & $2.461$ & $4.230-4.360$\\
\hline
\end{tabular}
\caption{\label{t:bes3} Publications by the BES-III collaborations listed in decreasing order of number of citations, with mention of datasets used for the analysis and the center-of-mass energy of the collisions. For some publications the exact integrated luminosity is not reported by the experiment in the article.}
\end{table*}
\normalsize

The BES-III experiment is a clear example of how the precise run schedule defines an experiment-wide utility function. BES-III studies electro-positron collisions in an energy range where tau physics, charm physics, and exotic hadron resonances may be studied in high detail, as are the properties of QCD at low energy. The number of interesting reactions, bound states, and phenomena that the experiment is sensitive to is very large, but each of these targets demands the collection of a different dataset. Wide-range energy scans allow the study of the $R$ evolution and cross-section dependencies on center-of-mass energy, as well as determine form factors in hadron production, while runs at the peak of resonances such as the $\psi(2S)$ enable precision measurements of charm states. Further, narrow scans at threshold are needed for measurements of the tau lepton mass, as well as for the estimate of hadron resonances parameters and branching fractions; while for searches of narrow resonances, fine-grained energy scans are needed where at each energy point are collected a sufficient amount of data to warrant observation of potential new states. In such a situation, the experiment must carefully design a run schedule which may optimize the scientific throughput by allocating days of running to different center-of-mass energy values. 

A look at the 20 most cited among the 659 publications produced by the BES-III collaboration to date (see Table~\ref{t:bes3}) shows how articles reporting the observation of new hadron states are among those of highest interest for the HEP community; they collectively brought 71.3\% of the citations of analysis articles ({\em i.e.}, the 20 listed except the first and third one, which deal with the detector and the future plan of the experiment, respectively) in the reported set. Indeed, in this energy regime hadronic physics is to date still concealing many ill-understood phenomena, and empirical classification remains a prime tool of investigation. A larger statistics of collisions at finely spaced energy points might allow BES-III to shed more light on the complex cross-section behaviour above the $\psi(2S)$ region --- a problem known as the ``Y'' problem; the scientific worth of data enabling other measurements of charmonium properties, form factors, and tau lepton mass is easier to assess.

\begin{table*}[h!]
\small
\begin{center}
\begin{tabular}{clrrr}
Energy (GeV) & Physics motivations & Final L ($fb^{-1}$) & Run days C & Run days U  \\
\hline
1.8-2.0   &  R values, nucleon cross sections      & 0.1 & 60 & 50  \\
2.0-3.1   & R values, cross sections& N/R & 250 &80 \\
3.686     & Light hadron and glueball charmonium decays &  $4.5$ & 150 &90 \\
3.770     & D meson decays & $20.0$ & 610 & 360   \\
4.180     & $D_s$ decay, XYZ states & $6.0$  & 140 & 50 \\
4.0-4.6 & XYZ, higher charmonia cross sections & $30.0$ & 770 &310 \\
4.6-4.9 & Charmed baryon and XYZ cross sections & $15.0$ & 1490 & 600 \\
4.74 & $\Sigma_c \bar{\Lambda_c}$ cross sections & $1.0$ & 100 & 40 \\
4.91 & $\Sigma_c \Sigma_c$ cross sections & $1.0$ & 120 &50 \\
4.95 & $\Xi_c$ decays & $1.0$ & 130 & 50 \\
\hline
\end{tabular}
\caption{\label{t:bes3goals} Declared goals of future BES-III studies of $e^+e^-$ collisions, collision energy, and required integrated luminosity and run time for the current (C) or an upgraded (U) BES-II detector; N/R stands for non reported. Excerpt from~\cite{bes3}. }
\end{center}
\end{table*}

The decision on how to divide the run time of the accelerator between fine-grained, wide-range energy scans and careful measurements at threshold or at the peak of known resonances is a complex one to make, as it involves a measure of arbitrariness. Yet the definition of a precise plan is still possible: indeed, in their 2020 study~\cite{bes3} the BES-III collaboration does not shy away from detailing the luminosity desiderata of BES-III for the near future, after appraising the various goals of the experiment. We report a concise version of a table (Table 7.1) presented in the Summary section of that document in Table~\ref{t:bes3goals}, to show how BES-III explicitly quotes the exact number of days to run at each energy point, separately for a current or upgraded version of the detector. We believe the table closely tracks the scientific value that the collaboration appraises to the measurements and searches that justify the collection of those datasets and is, therefore, a good illustration of how a multi-purpose HEP experiment can indeed formulate a global utility function.

\section{Case Study 2: Gamma-Ray Astrophysics
\label{s:swgo}}

Here we examine the case of a ground-based array of water Cherenkov detectors measuring high-energy gamma rays in the TeV--PeV range, in order to study how the different scientific goals of an astro-particle physics experiment may impose conflicting requirements on the detector geometry. We consider a number $N_{det}$ of idealized detector units placed on the ground at high altitude (4,800~m above sea level), corresponding to the conditions of the site (Pampa la Bola, in the Chilean Andes) chosen by the SWGO Collaboration~\cite{swgo} for a future detector of this kind. SWGO will detect radiation, electrons and muons of extensive atmospheric particle showers generated in the atmosphere by cosmic gamma-rays (the signal) and distinguish them from the large background from charged particles (mostly protons). A global goal of the experiment is to correctly tag and reconstruct the rare gamma-ray events in the haystack of cosmic rays: 1 every $10^{4}-10^5$. The particle shower hits the ground within an ellipse spanning thousands to millions of square meters around the shower development axis. The core science case of SWGO is quite extended, as it ranges from a study of the morphology of the steady PeV emission from the wide region of the galactic center, to the energy-dependent cosmic ray transport in galactic targets such as pulsar wind nebulae, to varying emission of active galactic nuclei, to the search of peculiar emission from dark matter dominated objects, {\em etcetera}~\cite{swgo}. The instrument's figure of merit is therefore defined by its energy and angular resolution, as well as by the gamma-ray acceptance and background discrimination capability; these ingredients play in differently for each of the scientific use cases mentioned above.

In order to meaningfully sample the area on the ground hit by shower particles, an array of about 6,000 water Cherenkov detectors is foreseen, with each detector composed by a water-filled tank with a diameter of a few meters. The most advantageous ground distribution of such an array is debatable: Is it better to strive for a homogeneous sampling, or to realize a decreasing fill factor from the center? Are radially symmetric layouts better than asymmetric ones? And does this depend on the specific science case considered? 

For the sake of studying the dependence of the instrument performance on the layout geometry, it is possible to abstract away from the details of the detection process, and assume that each unit measures with 100\% efficiency all secondary particles traversing it with an energy above some minimal threshold ({\em e.g.}, $E>10$~ MeV), with perfect discrimination between soft (electrons, positrons, gamma) and hard component (muons). Such a non-realistic ``asymptotic'' detection performance is purposely chosen to prevent the arbitrariness of a realistic, non-perfect detector performance (which depends on the detailed design of each detector) from affecting the appraisal of the relative merit of different configurations: by getting rid of ancillary confounders, we can better focus on the problem at hand, at least initially, to try and learn how geometry alone affects ultimate performance possibilities of the instrument. Any loose coupling of less-than-perfect efficiency and identification power with the details of the layout of detection units on the ground can be studied later as a refinement, as the detection technology of each detector unit can be fine-tuned at a second stage.

In order to study the extraction of meaningful physical measurements from cosmic showers, that include as we mentioned above a large background from primary protons along with the gamma-ray signal component, we use a closed-form parametrization of the radial distribution $dN/dR$ of the number of secondary particles of different kinds around the shower core, expressed as a function of primary particle identity (proton or gamma-ray), energy $E$, and angles of incidence (polar angle $\theta$ and azimuthal angle $\phi$), which we have developed to study the optimization of the SWGO array in~\cite{swgoopt}. Following that work, we consider the problem of defining an array of $N_{det}$ detector units laid on flat ground as a $2N_{det}-3$-parameter optimization problem, with the $i$-th detector contributing with the $x_i$ and $y_i$ coordinates of its center, and with three degrees of freedom being defined by the center of the array (set to $0,0$) and the choice of the $x$ axis direction. The problem can be simplified by considering ``macro-tank'' clusters of $M$ tightly packed detector units, which reduces the dimensionality by a corresponding factor; following~\cite{swgoopt}, $M$ is here chosen to be 19 (for a three-ring hexagonal cluster around a central tank at position $x_i$, $y_i$, with now $i=1,...,N_{det}/M$); a further reduction of dimensionality by a factor 3 is operated by enforcing 120-degree symmetry around the center of the array. Starting from an arbitrary default configuration, the array may be modified by letting the position of each triplet of macro-tanks change following the gradient of a utility function which we wish to maximize: \par

\begin{equation}
    x_i \to x_i + \eta_i \frac{dU}{dx_i} 
\qquad
    y_i \to y_i + \eta_i \frac{dU}{dy_i}
    \label{eq:updates}
\end{equation}

\noindent 
where the $\eta_i$ parameters are dynamically adjusted to dampen oscillating behavior and rewarding consistent movements of each detector in one specific direction on the plane, and where a vectorial average of gradients over the three macro-tanks laying in a symmetric position around the center is implicit. The utility function and its derivatives with respect to a change of each detector's position on the ground are computed using at each iteration a batch of simulated gamma and proton showers, whose parameters ($E$, $\theta$, $\phi$, and shower core position on the ground $X_0$, $Y_0$) are reconstructed with a likelihood maximization separately under each of the two primary hypotheses. The log-likelihood ratio for the two hypotheses serves as a test statistic to discriminate the signal component. For more details, we refer the interested reader to the stem article~\cite{swgoopt}.

\subsection{Choices of the Utility}

The detection and precise measurement of high-energy gamma rays opens the way to a number of studies of astrophysical sources and phenomena in the cosmos. Similarly to the case of a particle collider detector we discussed {\em supra}, it is hard to come up with a recipe that correctly appraises the relative value of a given precision in the determination of the gamma-ray flux in some specific energy range, the energy resolution, and the angular resolution. Further, the discrimination of proton from gamma primaries affects directly the overall performance of the instrument at the very high-end of the energy spectrum, where the gamma signal component is drowned in hadronic showers backgrounds that exceed it by four to five orders of magnitude.

We rewrite the utility function discussed in~\cite{swgoopt} below:\par
\begin{equation}
U_{1} = \lambda_{GF} U_{GF} + \lambda_{IR} U_{IR} + \lambda_{PR} U_{PR}
\label{eq:u1}
\end{equation}

\noindent
where the utility $U_1$ is composed of three partly independent contributions: 
$U_{GF}$, a term proportional to the precision in the determination of the gamma-ray flux for a given integration time of the observatory (defined as the estimated flux divided by its uncertainty); $U_{IR}$, a term inversely proportional to the expectation value of the uncertainty in the energy of detected gamma rays integrated over the full considered energy spectrum; and $U_{PR}$, a term inversely proportional to the expected uncertainty in the direction of incidence of the gamma rays. This definition allows us to exemplify the significant challenges posed by the optimization of the detector layout. To optimize the detector's positions in the plane, defined by their $x_i$, $y_i$ coordinates (with $i=1,...,N_{det}$, or if macro-units consisting of several detector units are used, $i=1,...,N_{det}/M$ with $M$ the number of units per macro-unit), we need to obtain the gradient updates of Eq.~\ref{eq:updates}. This involves computing the following expressions involving derivatives of the $U_1$ components: 

\begin{equation}
\frac{dU}{dx_i} = w_{GF} \lambda_{GF} \frac{dU_{GF}}{dx_i} + w_{IR} \lambda_{IR} \frac {dU_{IR}}{dx_i} + w_{PR} \lambda_{PR} \frac{dU_{PR}}{dx_i}
\end{equation}
\begin{equation}
    \frac{dU}{dy_i} = w_{GF} \lambda_{GF} \frac{dU_{GF}}{dy_i} + w_{IR} \lambda_{IR} \frac {dU_{IR}}{dy_i} + w_{PR} \lambda_{PR} \frac{dU_{PR}}{dy_i}.
\end{equation}

\noindent
Here, the introduced parameters $w_{GF}$, $w_{IR}$, and $w_{PR}$ are gradient-scaling factors introduced to balance the relative contributions of the gradients of the three utility components. While the coefficients $\lambda_{GF}, \lambda_{IR}, \lambda_{PR}$ reflect the scientific priorities of the experiment, the scaling factors $w$ need to be adjusted dynamically during optimization to account for the disparity in the magnitudes of the gradient components. This ensures that no single term dominates the updates to the detector positions, thereby improving convergence and maintaining a balanced optimization. The need for such scaling becomes apparent when considering the sensitivity of each term to detector movements. The $U_{GF}$ term, for example, can experience large changes due to the outward displacement of a single detector located on the array’s outer rim, which allows it to detect showers previously outside its range. In contrast, the $U_{IR}$ and $U_{PR}$ terms exhibit much smaller variations with individual detector movements, as their definitions involve averages over all detected showers, where only a subset of detectors contribute to the reconstruction of any given event.

Without the $w$ scaling factors, the gradient contributions from $U_{GF}$ often dominate, leading to updates that prioritize flux-related improvements at the expense of energy and pointing resolutions. This can result in a configuration where $U_{GF}$ increases significantly, but the overall utility $U_1$ may decrease due to a substantial degradation in $U_{IR}$ and $U_{PR}$. Such imbalances are further amplified by the collective movement of all detectors during a single gradient update, as the coupling between the large number of independent detector positions introduces additional dependencies that complicate the optimization landscape. The non-convex nature of the utility $U_1$, arising from the interplay of its components, further exacerbates this complexity, as it creates a landscape with potentially many local optima, saddle points, and flat regions. Consequently, the precise rule for dynamically varying the $w$ parameters may be difficult to define and is liable to affect unpredictably the capability of the gradient descent algorithm to converge to solutions at the Pareto front.

It must be noted that the model at the basis of the study of the detector optimization in~\cite{swgoopt} is a considerable simplification of the real physics scenario of the SWGO experiment, because of various approximations and assumptions that were taken in its construction. However, the conclusions we drew above on the complexity of the task are not affected by those simplifications. 

\subsection{An example}

\begin{figure}
\begin{center}
    \includegraphics[width=12cm]{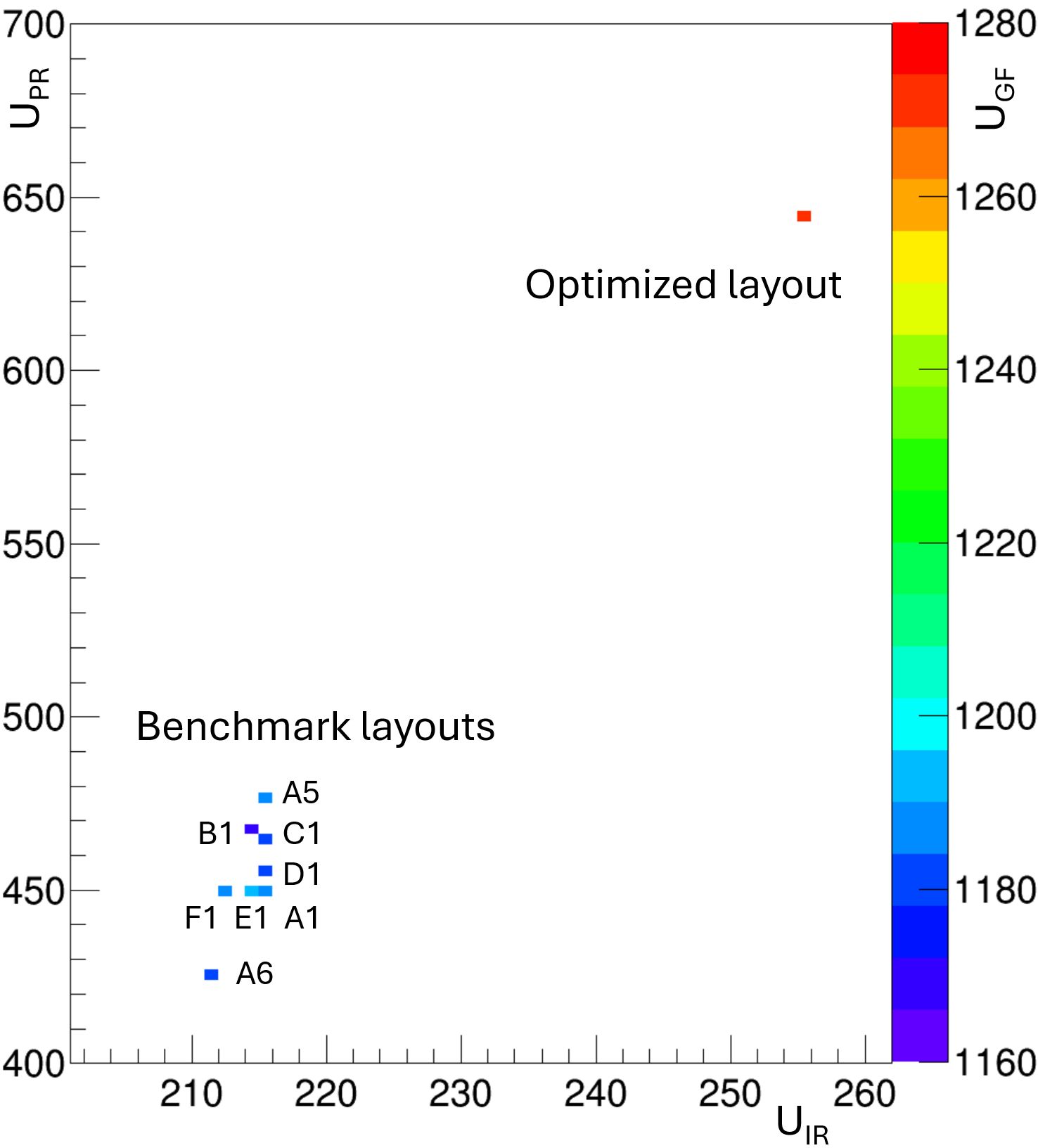}
    \caption{Value of the three components of the utility for 330 19-tank macro-units. The 8 benchmark layouts A1, A5, A6, B1, C1, D1, E1, F1~\cite{swgoopt} are compared to an optimized layout. See the text for detail.}
    \label{f:pareto}
\end{center}
\end{figure}

To illustrate how, in spite of the difficulties mentioned above, the search for an optimal point in the full-dimensional configuration space can be advantageous even in the presence of conflicting multi-target optimization criteria, we compare in Fig.~\ref{f:pareto} the three components of the utility function (Eq.~\ref{eq:u1}), for arrays of 330 19-unit macro-tanks laid down in configurations that mimic the A1, A5, A6, B1, C1, D1, E1, F1 SWGO benchmark layouts of 300-meter radius described in~\cite{swgoopt}. We use for this purpose batches of 2000 showers sampled from a uniform energy distribution in $[0.1-10]$ PeV. The figure shows how the eight benchmark configurations do not differ significantly in the value of the three utility components $U_{GF}$, $U_{IR}$, and $U_{PR}$; by only examining the utility values of those eight solutions one might be led to believe that they lay close to the Pareto front. However, a 1000-epoch gradient descent loop searching for the best configuration of the same number of units, initially deployed within a uniformly populated 300-meter-radius circle, and targeting the optimization of the sum $U_1^{*} = U_{IR}+U_{PR}$ alone, achieves quite significantly better performance than the highest achiever of the 8 benchmarks  on both components of the maximized utility ($(+19\pm 5)$\% on $U_{IR}$, and $(+35 \pm 4)$\% on $U_{PR}$), while also marginally improving (by $(6\pm4)$\%) the non-considered $U_{GF}$ component. Further examples are discussed in~\cite{swgoopt}.

\subsection{Energy dependence}

We now return to the full definition of the $U_1$ utility, and consider how the optimal definition of the layout of detectors may be strongly dependent on the goals of the experiment, by simply comparing optimized layouts that considered atmospheric showers sampled from different energy distributions: the assumed relative frequency of higher- or lower-energy showers in the training datasets acts as a weight in determining the relative value of showers of different energy in the utility. We control the distribution of showers energy by a coefficient in the power law defining their density function:\par
\begin{equation}
    f(E) = k e^{-SE}
\end{equation}
\noindent
where the constant $k$ is a normalization factor and where energy $E$ is considered, in the considered setup, to vary from $100$~TeV to $10$~PeV. In practice, by changing the $S$ parameter from 1.0 to -1.0, we effectively mimic an emphasis of the experiment on the physics of showers in the few hundred TeV range, to an emphasis in the several PeV range. By running optimization jobs that rely on data sampled from those different distributions we may therefore observe what is the corresponding effect on the layout; further, we may then compare the performance of the optimized layouts when tasked with reconstructing showers following a different energy distribution, to verify what price is paid in choosing layouts optimized with different energy priors.

\begin{figure*}[h!]
\includegraphics[width=0.99\linewidth]{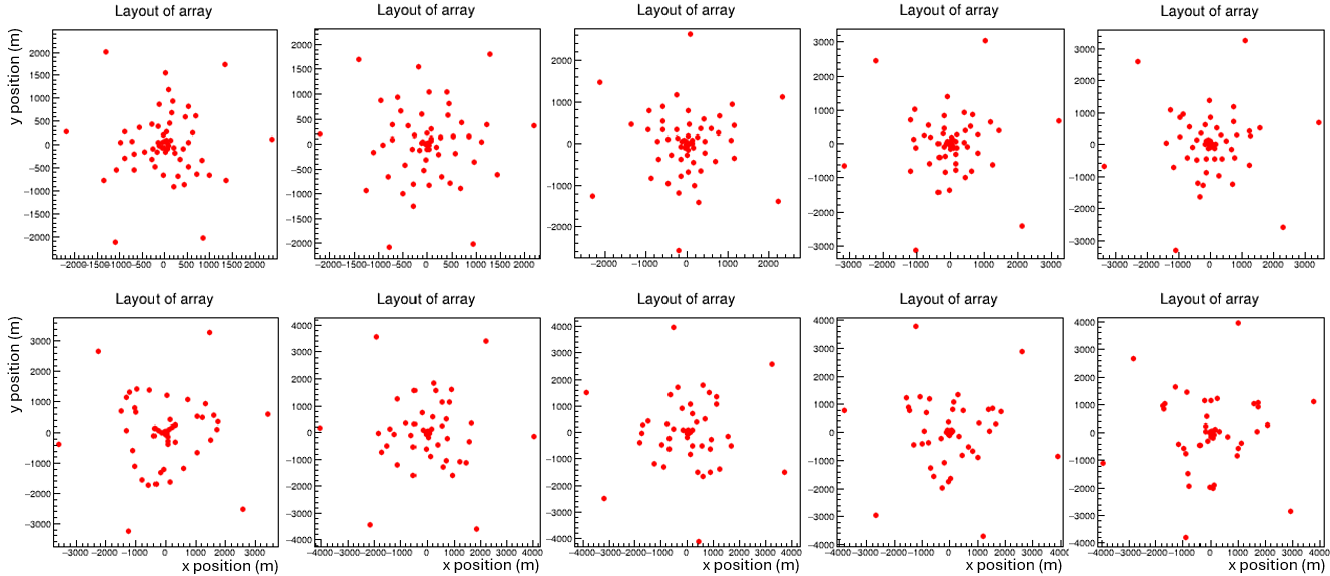}
\caption{Distribution on the ground of 60 macro-units resulting from the optimization of the utility function described in the text, for gamma rays following a power law with coefficients $S$ varying from +1.0 (top left) to +0.2 (top right) and from 0.0 (bottom left) to -0.8 (bottom right). See the text for more detail.}
\label{f:layouts}
\end{figure*}

Figure~\ref{f:layouts} shows 10 optimized configurations of 1140 detection units, where the units are joined into 60 19-unit aggregates whose position is separately optimized. As mentioned {\em supra}, a triangular symmetry is enforced by the algorithm to reduce the dimensionality of the problem, such that the pattern on the ground is made up by three equal 120-degree sections. All in all, this corresponds to a 37-dimensional design problem which does not pose very significant computing demands; 1000-epoch updates are performed on the detector layouts, with a starting configuration packed in a circle around the origin.

We may observe that the overall appearance of the shown configurations is similar, with the strikingly common feature of six macro-tanks deployed at large distance from the center in a hexagonal arrangement, and a more complex central arrangement. The outer macro-tanks help in the reconstruction of angular and energy parameters of the most extended, highest-energy showers while guaranteeing that they get accepted by the trigger selection, which in these simulations filters for reconstruction showers that yield a signal in $N \geq 50$ detectors\footnote{Given that each macro-unit is an aggregate of 19 detector tanks, in principle a signal in all detectors of a single macro-unit might suffice to meet the trigger criterion, but the reconstruction would then return very ill-measured shower parameters.}. Other than that, there are important differences: in particular, optimization runs targeting an energy spectrum with $S>0$ produce less extended layouts than ones using $S<0$, as higher-energy showers more frequent in data used by the latter have a larger footprint on the ground, and thus benefit from wider arrays.

\begin{figure*}[h!]
\includegraphics[width=0.99\linewidth]{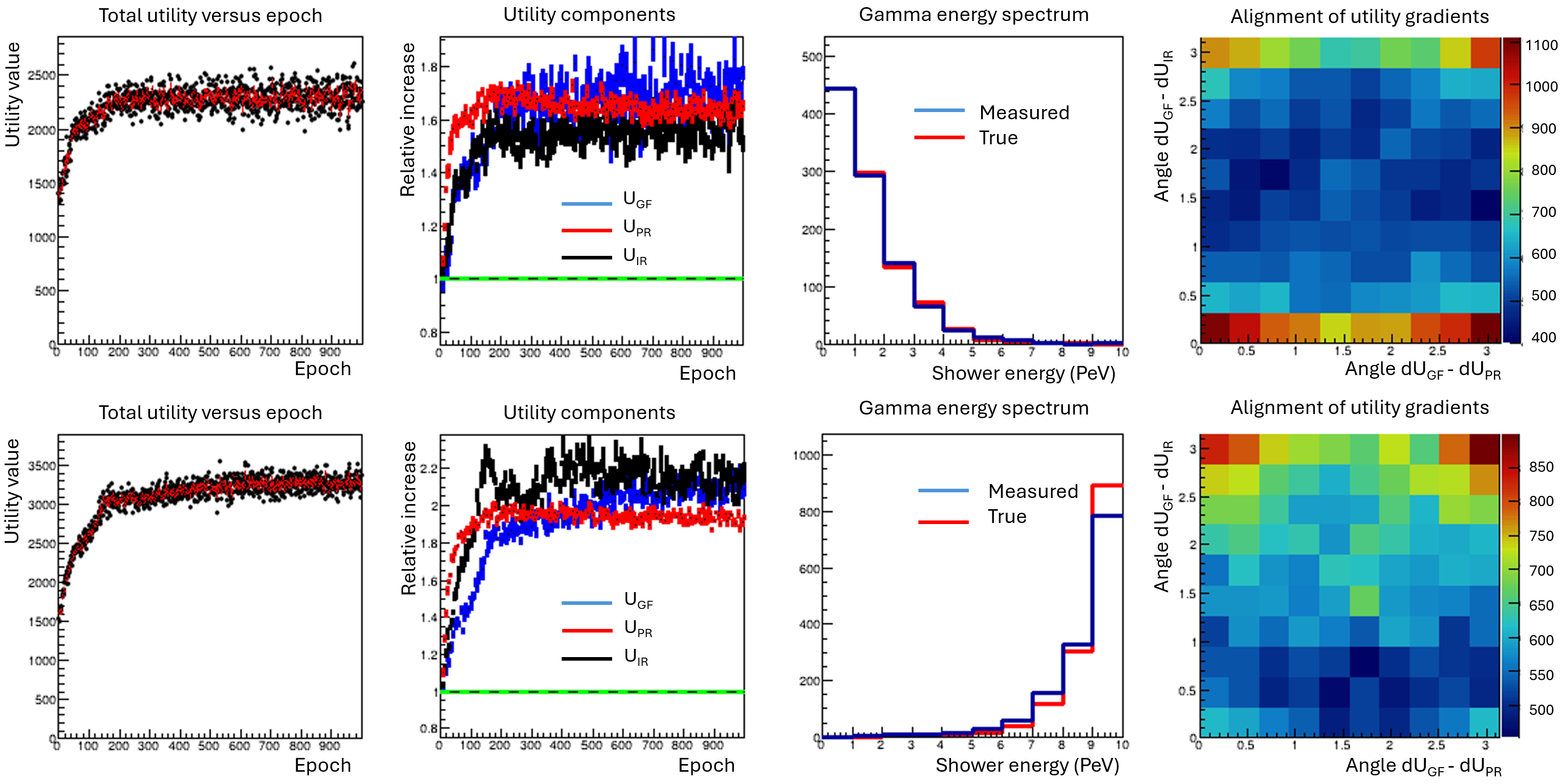}
\caption{\em From left to right, the graphs show: evolution of the total utility $U_1$ as a function of epoch number; evolution of the three components of the utility versus epoch; generated and reconstructed spectrum of gamma-ray energies; and angle between the gradient vector of pairs of utility components during the optimization. The top line refers to the optimization based on a gamma-ray spectrum with $S=1.0$; the bottom line refers to the optimization based on a gamma-ray spectrum with $S=-1.0$. See the text for more details.}
\label{f:detail}
\end{figure*}

If we examine the graphs shown in Fig.~\ref{f:detail}, we see an example of the complexity of an optimization task with a multi-target utility. While the relative multipliers $w$ of the three utility gradients are automatically equalized during the optimization, some of the utility components may still not increase as much as others (second panel from the left in the two rows). Also, the rightmost panels of the figure show how the three two-dimensional gradient vectors in the detector plane are usually not aligned with one another: for the high-energy $S=1.0$ case the gradient of the flux utility $U_{GF}$ is mostly pointing in the opposite direction of the gradient of the integrated energy resolution piece $U_{IR}$; the opposite behavior is apparent for the $S=-1.0$ case. A more mixed behavior is seen for the alignment between the gradients of the $U_{GF}$ and pointing resolution utility component $U_{PR}$.

\begin{table}[h!]
\begin{centering}
    \begin{tabular}{rll}
    $S_{opt}$ & $U_{S=1.0}$ & $U_{S=-1.0}$ \\ 
    \hline
1.0      & 2360$\pm$21 & 2880$\pm7$  \\
0.8      & 2384$\pm9 $ & 3072$\pm8$ \\
0.6      & 2368$\pm10$ & 2901$\pm7$ \\
0.4      & 2358$\pm10$ & 2979$\pm10$ \\
0.2      & 2371$\pm11$ & 3070$\pm9$ \\
0.0      & 2220$\pm13$ & 3153$\pm9$ \\
-0.2     & 2255$\pm11$ & 3226$\pm10$ \\
-0.4     & 2271$\pm12$ & 3187$\pm9$ \\
-0.6     & 2247$\pm12$ & 3152$\pm10$ \\
-0.8     & 2098$\pm14$ & 3049$\pm9$ \\
-1.0     & 2234$\pm12$ & 3239$\pm$49 \\

\end{tabular}
\caption{\label{t:eslopevar} Comparison of the total utility $U_1$ estimated for 11 different layouts (each previously optimized based on gamma rays with energy spectra corresponding to 11 values of $S_{opt}$ from +1.0 to -1.0) using fluxes of low-energy-rich ($S=1.0$) or high-energy-rich ($S=-1.0$) gamma rays. Utility values should therefore be compared only within columns. }
\end{centering}
\end{table}

If we compare the utility values corresponding to gamma-ray spectra of low-energy ($S=1.0$) or high-energy ($S=-1.0$) for the 11 layouts optimized with fluxes of gamma rays generated with spectra corresponding to values of $S$ varying from +1.0 to -1.0 (see Table~\ref{t:eslopevar}), we see how the choice of a target energy spectrum at optimization stage affects the potential of the experiment in the study of gamma rays of different energy: an array optimized for $S=+1$ will lose 8-10\% of its effectiveness when studying gamma rays coming from a harder spectrum, and an array optimized for $S=-1.0$ will similarly lose about 10\% in terms of its utility. 

It should be clear how the typical staged approach of construction of extended arrays such as SWGO --- that may decide to build first an array with a limited number of detection units, and later add more units if more funding becomes available --- may pose a challenge: {\em e.g.}, the collaboration might decide to target a lower energy range in the first stage, and thus build the array in a configuration optimized for that purpose; when adding more detectors to then target a higher energy, the first part of the detector remains however fixed (as changing the position of the initial units would constitute a very significant additional cost); the resulting extended array cannot then reach the same performance it would have if the position of all units had been optimized for the higher energy range in the first place.

\subsection{A word on flexible configurations }

It is useful to mention here that in some cases it may be possible to design an experiment in a way that allows for seamless reconfiguration of its elements to fulfill different tasks, and reach optimality for each. That is the case, {\em e.g.}, of the ALMA array of radio-telescopes in northern Chile~\cite{ALMA}. To achieve this kind of ultimate flexibility, the individual elements of the array are built such that they are movable with custom trucks over the high-altitude plain where the array is deployed. ALMA can thus arrange its radio-telescopes in a tight cluster to achieve high performance for large-aperture surveys, or over a wide area to maximize its angular resolution. The spectacular scientific output of the experiment is a testament to the success of this strategy.

It is necessary to point out that reconfigurable arrangements impose significant constraints on the weight and dimensions of the re-deployable elements, and that in particular, they cause a significant budget overhead over that of a static array. For an array such as SWGO, where each detector unit needs water, power, and signal cable connections to a central hub and counting house, and when the number of units is eventually foreseen to reach several thousand, the option is completely impractical. One possible workaround in this context is to construct and position a larger number of detector tanks on the ground than can initially be instrumented and operated. In the first phase, only a subset of these tanks would be filled with water and equipped with detection instruments, based on an optimized configuration for the initial scientific goals. In a second phase, water and detection instruments could be relocated to some of the initially inactive tanks, allowing the array configuration to be adapted to optimally study a different set of goals; or, if more funding became available, all tanks could be made operational. In any case, when such solutions can be implemented, budget remains the main driver of a final decision.

\section{Conclusions \label{s:conclusions}} 

Machine learning tools today offer the possibility to explore the very large space of design configurations for fundamental science experiments in a continuous way, through the generation of differentiable surrogate models of the stochastic elements that may be present in the data generation processes. Such explorations may identify groundbreaking solutions by exploiting the subtle correlations and interplay of the many construction parameters of particle detectors and similar apparatus. In such a situation, the exact definition of an experiment-wide utility function, which may summarize the relative value of the various scientific goals of an experiment, becomes a crucial ingredient in the way of an end-to-end optimization of its hardware and software.

In this work, we have argued that the definition of a global utility function is not only possible, but implicitly done already in many experimental situations, and therefore it should not be considered a significant hindrance in the way of global optimization programs. The case of the BES-III experiment mentioned in Sec.~\ref{s:bes3} is a glaring demonstration of the fact that, when it is important for a large scientific collaboration to quantitatively define the relative worth of different scientific goals, the corresponding decisions are promptly and precisely taken. 

The conflict between different targets of an exploratory instrument such as a ground-based array for cosmic ray studies has been discussed in Sec.~\ref{s:swgo}. In that case, using an optimization program developed for the study of the layout of the SWGO observatory, we showed in practice how the design can be modified by changes in the preferred energy range of studied atmospheric showers. When possible, configurations that can be modified after construction would allow the experiment to retain optimality for a variable set of scientific goals during its lifetime.

\end{document}